\documentclass[fleqn,12pt,twoside]{article}
\usepackage[headings]{espcrc1}

\readRCS
$Id: espcrc1.tex,v 1.2 2004/02/24 11:22:11 spepping Exp $
\usepackage{graphicx}
\usepackage[figuresright]{rotating}

\newcommand{\AmS}{{\protect\the\textfont2
  A\kern-.1667em\lower.5ex\hbox{M}\kern-.125emS}}
\newcommand{ \rts }{$\sqrt{s_{_{\rm NN}}}$}
 
\def \auau  {Au+Au}
\def \dau   {$d$+Au}

\def \pt    {$p_T$}

\def \d0    {$D^0$}

\title{A Heavy--Flavor Tracker for STAR}

\author{K. Schweda for the STAR Collaboration
\address{Lawrence Berkeley National Laboratory, \\
One Cyclotron Rd MS70R0319, Berkeley, CA 94720, USA}
\thanks{Present address:
        University of Heidelberg, Physikalisches Institut, \newline  
        Philosophenweg 12, 69120 Heidelberg, Germany.}
\thanks{For the full list of STAR authors and acknowledgments, see appendix 'Collaborations' of this volume.}}
       
\runauthor{K. Schweda}

\begin{document}

% typeset front matter
\maketitle

\begin{abstract}
We propose to construct 
a heavy flavor tracker for the STAR experiment at RHIC 
in order to measure the elliptic flow of charmed hadrons 
in the low \pt region and identify
B-meson contributions in the region \pt\ $>$ 4 GeV/c. 
In this talk, we will present the design of the detector in-depth 
and its expected performance as studied in detailed simulations 
and analytic calculations. 
Physics potentials of the detector will also be discussed.

\end{abstract}

\section{\it Introduction}
Elliptic flow measurements have demonstrated that partonic collectivity,
collective flow of partons, has been developed in 200~GeV Au+Au 
collisions at RHIC~\cite{pc}. To pin down the partonic EOS of matter produced
at RHIC, one must 
address the status of thermalization in such collisions. 
Since the masses of heavy-flavor quarks, e.g. charm quarks, 
are much larger than the maximum possible excitation of the system 
created in the collision, heavy-flavor collective motion could be 
used to indicate the thermalization of light flavors 
($u,d,s$). 

The development of collectivity 
at the partonic level (among quarks and gluons) and the degree 
of thermalization are closely related to the equation of state 
of partonic matter: Re-scattering among constituents and the density profile lead to the 
development of collective flow. In case of sufficient re-scattering, 
the system might be able to reach local thermal equilibrium. 

Heavy-flavor quarks are special probes because of their heavy mass. 
If chiral symmetry is restored in a QGP, light quarks obtain their small 
current masses. On the other hand, heavy quarks get almost all their mass 
from their coupling to the Higgs field~\cite{Mueller}. 
Thus, heavy quarks  stay heavy -– even in a QGP. The observation 
of heavy-quark collective flow indicates multiple interactions among partons. 
This would suggest that light quarks are thermalized.
Here, the heavy-flavor transverse elliptic flow is an especially promising early stage observable,
while transverse radial flow might be cumulated throughout the whole collision history.

%
% figure 1
\begin{figure}[t]
\begin{center}
\vspace{-0.6cm}
\begin{minipage}{0.445\textwidth}
\includegraphics[width=0.99\textwidth]{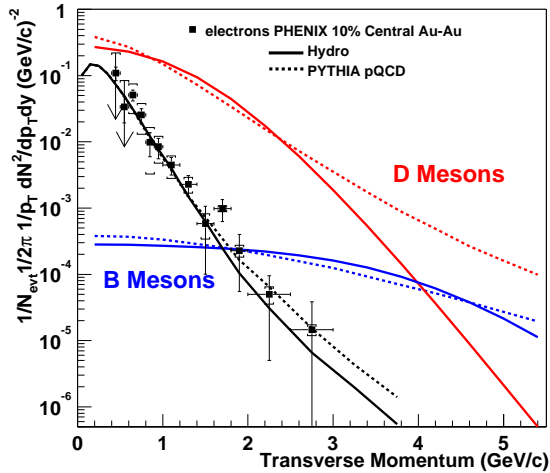}
\end{minipage}
%\begin{minipage}{0.49\textwidth}
%\vspace{-2mm}
%\includegraphics[width=0.99\textwidth]{D0_RAA2.eps}
%\end{minipage}
\vspace{-5mm}
\caption{Invariant yield of non-photonic 
single electrons from central \auau\ collisions at \rts = 200~GeV~\cite{PHENIX_e}. The dashed curves show results from pQCD calculations
(zero heavy-quark flow) for $D, B$ mesons and the combined resulting decay electrons. The solid curves represent results from
hydrodynamical model calculations (full heavy-quark flow). This figure has been taken from~\cite{Kelly}. 
%Right: Modification of the \d0 \ meson spectrum as a function of transverse momentum for 
%flow velocities $\langle\beta\rangle$ = 0.1, 0.3 and 0.4 (dotted, dashed and solid line) relative to 
%a \dau\ reference~\cite{STAR_e}.
}
\label{fig1a}
\end{center}
\vspace{-1.5cm}
\end{figure}
First results on heavy-flavor production at RHIC have been reported from
observing electrons stemming from the decay of heavy-flavor \mbox{quarks~\cite{PHENIX_e,STAR_e}}. 
However, due to the decay kinematics, important information on heavy-flavor dynamics is smeared \mbox{out~\cite{Kelly,Xin}}.
This is demonstrated in Fig.~2. The data points show the invariant yield of non-photonic 
single electrons from central \auau\ collisions at \rts = 200~GeV~\cite{PHENIX_e}. The dashed curves show results from pQCD calculations
(zero heavy-quark flow) for $D, B$ mesons and the combined resulting decay electrons. The solid curves 
represent results from hydrodynamically inspired model calculations (full heavy-quark flow). 
Both extreme dynamical scenarios reproduce the measured electron spectra.

At this conference, recent results on electron $R_{AA}$ have triggered lots of exciting
discussions. It seems that we do not fully understand the underlying mechanism of heavy-flavor interaction
with the dense medium. At higher \pt, therefore, it is also important to measure distributions
from directly reconstructed $D$-mesons in order to isolate the bottom contributions in collisions at RHIC.

%The modification of the \d0 \ spectrum is further quantified by taking the ratio $R_{AA}$
%of the spectra expected from \auau\ collisions (flow) relative to \dau\ collisions (non-flow).
%Here we parametrized experimental results for $D$ mesons, 
%directly reconstructed through the invariant mass of their decay
%daughters from \dau\ collisions as measured in STAR~\cite{STAR_e}, by a pQCD-inspired power-law fit. 
%For \auau\ collisions, we followed a similar hydro-dynamically inspired approach as above,
%with a kinetic freeze-out temperature $T_{\rm fo}$ = 160 MeV and an average flow velocity $\langle\beta\rangle$ = 0.4 
%(in units of speed of light). A radially linear flow profile was assumed. 
%The right panel in Fig.~1 shows the modification of the $D^0$ spectrum 
%as a function of transverse momentum for three different
%flow velocities. The modification is in the order of 30-50\% with the maximum moving to
%larger momentum with increasing flow velocity.

STAR has measured $D$-mesons in \dau\ and \auau\ collisions by direct reconstruction through
the invariant mass of decay-daughter candidates~\cite{STAR_e,Haibin}.
Due to the large multiplicities of $\pi$,K,p and the rather small production cross section
for charm-hadrons, the combinatorial background in the invariant mass distribution
is roughly 1000 times larger than the signal~\cite{Haibin}. Extending particle identification by time
of flight information will improve the statistical significance by a factor of five.
This large combinatorial background leads to systematic uncertainties of extracted charm-hadron 
yields in the order of 30\%. On the other hand, elliptic flow modulates particle yields with respect to
the reaction plane in the order of 10\%. 
To overcome these large systematic uncertainties and make precise heavy-flavor 
elliptic flow measurements feasible,
we propose to upgrade STAR with $\mu$-vertex capabilities to identify heavy-flavor hadrons through
their displaced decay vertex~\cite{HFT}.    

\section{\it Mechanical Setup}
A perspective view of the mechanical setup of the proposed Heavy-Flavor Tracker~(HFT) is shown in Fig.~2. 
It sits inside the STAR Time Projection Chamber. The length is 20~cm, covering $\pm$1.1 units 
in pseudo-rapidity. It has two tracking layers composed
of 2x2~cm$^2$ monolithic CMOS sensors with 30x30~$\mu$m square pixels at radii 
1.5~cm and 5.0~cm, covering full azimuth. The pixel granularity is 100k/cm$^2$ or 100~M pixels in total.
Several proto-types of these sensors, called MIMOSA, have been built by the IReS group in Strasbourg~\cite{Winter}.
A sensor efficiency of better than 99\% has been achieved.
These sensors are thinned down to 50$\mu$m.
The total material budget of 0.36\% radiation length per layer 
includes the active sensors, readout chip, cabling and support structure.
This minimizes distortion of charged particles through
multiple Coulomb scattering. Low power consumption of less than 100~mW/cm$^2$ allows for air cooling of the detector. 
A single-sided mounting support
will hold the HFT on a finger-like structure. This mounting support fits into the present STAR detector
mechanical setup enabling reproducible detector alignment to better than 10~$\mu$m. The mechanical
stability was tested on an optical setup applying interferometry. The stiffness and bending characteristics
meet our specifications. E.g. vibrations induced by air flow for cooling introduced an uncertainty in
the position location of less than 2~$\mu$m. 
\begin{figure}[bh]
\vspace{-5mm}
\begin{minipage}{0.46\textwidth}
\includegraphics[width=.9\textwidth]{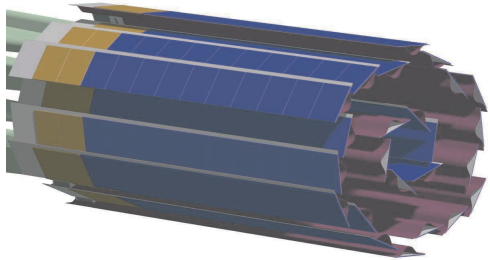}
\end{minipage}\hspace{-0.0cm}
\begin{minipage}{0.49\textwidth}\vspace{-8mm}
\caption{Perspective view of the STAR Heavy-Flavor Tracker. The length of the ladders is 20 cm in beam-direction,
the two layer radii are 1.5~cm and 5.0~cm, covering full azimuth. The pixel granularity 
is 100k/cm$^2$ or 100~M pixels in total. On the left, the single-sided finger-like support structure is partially visible.}
\end{minipage}
\label{fig2}
\end{figure}
\vspace{-15mm}
\section{\it Tracking Simulations}

The expected performance of the HFT has been studied in detailed simulations.
We used a Monte Carlo event generator, parametrizing experimental particle distributions from
\auau\ collisions at \rts = 200~GeV. Charged particles were propagated through the full detector geometry
by means of GEANT 3.21. The generated information was then fed through realistic response simulators 
and a full tracking algorithm was applied. For the HFT, Monte Carlo hits were smeared by a Gaussian function
of width $\sigma$=6~$\mu$m in $y$ and $z$-direction to account for the finite pixel-hit resolution.
The decay channel $D^0 \rightarrow$ K + $\pi$ (BR= 3.8\%, $c\tau=123\mu$m) was studied. 
Decay and topological cuts were optimized~\cite{alex}
using the minimization package MINUIT. A $D^0$ signal with a statistical
significance of 3-$\sigma$ is observed with 8k central collisions.

Figure~3 shows results on the elliptic flow of \d0 -mesons from model predictions~\cite{Molnar} 
assuming full charm quark flow (solid line) and no charm quark flow (dashed line).
The expected statistical uncertainty after one year of data
taking, assuming 50M events, is shown on the top line. 
The uncertainties are largest at low momentum due to the rather small \d0 \ reconstruction efficiency, 
they reach a minimum around 2 GeV/c, and then increase due to the exponentially falling 
\d0 \ yield at larger momentum. 
The differences in the predictions for both extreme scenarios are in the order of a factor two, 
while the projected statistical uncertainties are expected to be smaller than 10\% in the momentum region 1-3GeV/c. 
Hence, with one year of data taking, the question of charmed quark flow can be fully addressed.

We also studied another charm-hadron decay, i.e. $D_s^+ \rightarrow \phi + \pi$ (BR= 3.6\%, $c\tau=150\mu$m), 
with $\phi \rightarrow K^+ + K^-$ (BR= 49.2\%).
Here, the momentum coverage is 1.0$<$\pt$<$3.0 GeV/c, covering 60\% of the integrated yield. The ratio
\d0 / $D_s^+$ is especially sensitive to different charm-quark hadronization scenarios, giving further insight into 
charm-quark dynamics~\cite{Andronic}.
%
% figure 3
\begin{figure}[thb]
\vspace{-7mm}
\begin{minipage}{0.49\textwidth}
\includegraphics[width=0.95\textwidth]{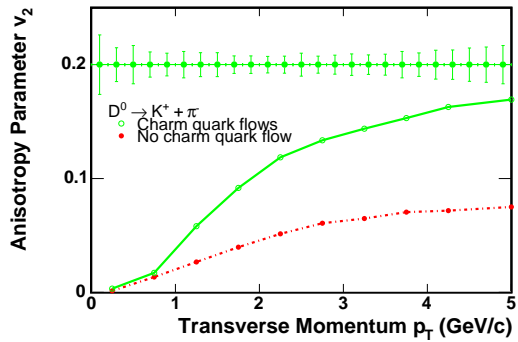}
\end{minipage}\hspace{-0.0cm}
\begin{minipage}{0.49\textwidth}\vspace{-8mm}
\caption{Results on the elliptic flow of \d0 -mesons from model predictions~\cite{Molnar} 
assuming full charm quark flow (solid line) and no charm quark flow (dashed line).
The expected statistical uncertainty after one year of data
taking, assuming 50M events, are shown by solid circles.} 
\end{minipage}
\label{fig3}
\end{figure}
\vspace{-7mm}
\section{\it Summary}
The precise measurement of heavy-flavor hadron elliptic flow, spectra and yield will help address the
exciting topic of light-quark thermalization in high-energy nuclear collisions. These measurements require
large momentum coverage to low \pt\ at small background. The proposed Heavy-flavor Tracker for STAR 
applies active pixel sensor technology with a position resolution better than 10~$\mu$m, at a low material budget
(0.36\% radiation length per ladder) and high mechanical stability. Precise measurements on
heavy-flavor production will be feasible. Our goal is to complete construction and installation of the HFT
before the next long \auau\ run at RHIC.    

\end{document}